\documentclass[twocolumn,showpacs,amsmath,amssymb]{revtex4}

\usepackage{graphicx} 
\usepackage{dcolumn}  
\usepackage{bm}       

\def\be{\begin{eqnarray}}
\def\ee{\end{eqnarray}}
\def\bse{\begin{subequations}}
\def\ese{\end{subequations}}
\def\l{\langle}
\def\r{\rangle}
\def\a{\alpha}
\def\b{\beta}
\def\g{\gamma}
\def\p{\prime}
\def\ra{\rightarrow}
\def\lra{\longrightarrow}

\def\ap{\alpha^{\prime}}
\def\bp{\beta^{\prime}}

\newcommand{\ket}[1]{\left| {#1} \right>}
\newcommand{\bra}[1]{\left< {#1} \right|}

\newcommand{\C}{^C\!}
\newcommand{\cc}{^{CC}\!}

\begin{document}

\title{Further quantum-gate methods using selective displacement 
of trapped ions}

\author{Marek \v{S}a\v{s}ura and Andrew M. Steane}

\affiliation{Clarendon Laboratory, Department of Physics,
University of Oxford, Parks Road, Oxford OX1 3PU, United Kingdom}

\date{\today}

\begin{abstract}
We consider quantum gates for trapped ions using state-selective
displacement of the ions. 
We generalise earlier work in order to treat arbitrary separations between
the traps. This requires the impact of anharmonicity arising from the Coulomb 
interaction to be estimated. We show that its effects are always small enough
to allow high fidelity. In particular, the method can be applied to two
ions in the same trap. We also show that gates between non-neighbour ions,
and hence a Toffoli (three-qubit controlled-NOT) gate, can be achieved.
We discuss how the gate can be applied to logical qubits encoded in
the decoherence-free-subspace $\{ \ket{01},\, \ket{10}\}$,
where each pair of ions stores a single qubit. 
We also suggest alternatives
to the spin-echo method to suppress unwanted terms in the evolution.
\end{abstract}


\maketitle

\section{Introduction}

This paper discusses the implementation of quantum logic gates
in trapped ions, using a mechanism introduced
by Cirac and Zoller in Ref.~\cite{ions1},
in which the ions are displaced by a state-selective
force, and acquire state-dependent phases from their Coulomb
interaction. In a previous paper~\cite{ions3} 
we discussed in detail the realization of a two-qubit phase gate
\be
\label{first}
^{C}\!P(\theta)\equiv 
\ket{00}\bra{00}+\ket{01}\bra{01}+\ket{10}\bra{10}
+e^{i\theta}\ket{11}\bra{11}\,,\nonumber\\
\ee
implemented between two ions confined in separate traps. 
When $\theta=\pi$ this is the gate $^{C}\!Z$ which is equivalent 
to the controlled-NOT gate ($^{C}\!X$) up to single-qubit 
Hadamard rotations~$H$, where $X$
and $Z$ are the Pauli $\sigma_X$ and $\sigma_Z$ matrices.
In this paper we extend the scheme of the two-qubit phase gate 
in several ways.

First, we present some methods relevant to the two-qubit phase gate. These
are the use of different laser frequencies and/or switching force
directions to eliminate undesired rotations, 
and the application of the gate to manipulate logical information
encoded in a simple decoherence-free subspace. 

Next, we consider the three-qubit phase gate $\cc Z$ 
(controlled-controlled-$Z$ gate) whose effect is
\be
\label{3phase}
|\a\b\g\r\lra (-1)^{\a\b\g}|\a\b\g\r\,,
\ee
where $\alpha, \beta, \gamma \in \{0,1\}$.
This gate is equivalent to the Toffoli gate $\cc X$
(controlled-controlled-NOT)
in the sense that it suffices to add two single-qubit Hadamard
rotations to obtain the Toffoli gate. The combination of $\cc Z$
with the Clifford group (generated by $Z, H, ^C\!X$)
is a universal set for quantum computation.
In this respect $\cc Z$ is similar to $^C\!P(\pi/2)$. 
Both $\cc Z$ and $^C\!P(\pi/2)$ are used in networks 
to achieve fault-tolerant implementation of a universal set of gates on
qubits encoded in quantum error correcting codes~\cite{chuang, steane04}. 
They are therefore
desirable features of any quantum computing system.
We study the problem of how to use the state-selective displacement of
trapped ions (the ``pushing'' method) 
to achieve the transformation in Eq.~(\ref{3phase})
with three ions in adjacent microtraps. Our solution involves
a standard decomposition of $\cc Z$ into five two-qubit gates
(two controlled-NOT and three controlled-phase $^C\!P(\theta)$). 
In this decomposition,
one of the two-qubit gates is between non-neighbouring qubits.
The interesting feature of the pushing method under discussion
is that this gate between non-adjacent ions can be achieved without
the need to rearrange the ions or swap the quantum information
between them.

Finally, we extend the discussion of two-ion gates to the case where
the ions are much closer together than was previously assumed. In
this regime the Coulomb interaction introduces non-negligible
anharmonicity in the confining potential. We quantify this effect
and show how the anharmonicity influences
the sensitivity of the gate to
thermal motion of the ions. This more general treatment includes 
the case where two ions are confined in the same trap. We find that the
effect of anharmonicity remains small even in this limit, and therefore
the gate remains comparatively insensitive to thermal motion.

\section{Two-qubit phase gate}
\label{two}

In this section we briefly discuss the two-qubit phase gate
considered in Refs.~\cite{ions1, ions2, ions3},
in order to give the background to the rest of the paper.

Let us consider a system of two interacting qubits. Suppose the
evolution is of the form
\be
\label{two1}
|\a\b\r\stackrel{G}{\ \lra\ } e^{i\Theta_{\a\b}}|\a\b\r\,,
\ee
where $\ket{\alpha \beta}$ are states in the
computational basis $\{|00\r,
|01\r, |10\r, |11\r\}$. The phases $\Theta_{\a\b}$ have
values arising from the internal energy of each qubit,
and the interaction energy. We wish to obtain a
two-qubit phase gate, which produces an overall phase $\vartheta$
if and only if both qubits are in the logical state $|1\r$, that
is 
\bse 
\label{two2}
\be
|\a\b\r & \lra & |\a\b\r\,,\quad \a\b=0\,,\\[1mm]
|\a\b\r & \lra & e^{i\vartheta}|\a\b\r\,,\quad \a\b=1\, .
\ee
\ese
In order to obtain this
we next apply local operations (single-qubit rotations) on both qubits
\be
\label{two3}
S=S_1\otimes S_2\,,
\ee
where we will assume
\bse
\label{two4}
\be
S_1 &=& |0\r_1\l 0|\,e^{iA_0}+|1\r_1\l 1|\,e^{iA_1}\,,\\[1mm]
S_2 &=& |0\r_2\l 0|\,e^{iB_0}+|1\r_2\l 1|\,e^{iB_1}\,.
\ee
\ese
The sequence $SG$ corresponds to the phase gate~(\ref{two2}) 
if the following set of algebraic equations is satisfied
\bse
\label{two5}
\be
A_0+B_0+\Theta_{00} &=& 0\,,\\
A_0+B_1+\Theta_{01} &=& 0\,,\\
A_1+B_0+\Theta_{10} &=& 0\,,
\ee
\ese
where the overall phase of the gate is given by
\be
\label{two6}
\vartheta=A_1+B_1+\Theta_{11}\,.
\ee
Eqs.~(\ref{two5}) are three independent equations
in four independent parameters $(A_0,A_1, B_0,B_1)$. 
This means we can choose one parameter. 
Let us choose $A_0=-\Theta_{00}/2$.
Then, using Eqs.~(\ref{two5}), we get for the other three parameters
\bse
\label{two7}
\be
B_0 &=& -\Theta_{00}/2\,,\\
A_1 &=& -\Theta_{10}+\Theta_{00}/2\,,\\
B_1 &=& -\Theta_{01}+\Theta_{00}/2\,,
\ee
\ese
and the overall phase reads
\be
\label{two8}
\vartheta=\Theta_{11}-\Theta_{10}-\Theta_{01}+\Theta_{00}\,.
\ee
When $\vartheta=\pi$ we have the quantum logic gate $\C Z$ which is
equivalent to the controlled-NOT gate $\C X$.

Next we briefly discuss the spin-echo or $\pi$-pulse method which can be
used to suppress an unwanted term in the evolution and thus 
improve the fidelity (precision) of the gate.
This method relies on the assumption that we generate the same 
unwanted term in two successive operations $G$. 
A~pair of $\pi$~pulses on qubits 1 and 2 produces the effect
\be
\label{pi1}
R=R_1\otimes R_2\,,
\ee
where
\be
\label{pi2}
R_i=|0\r_i\l 1|+|1\r_i\l 0|\,.
\ee
We replace the original 
gate sequence $SG$ by $S'(RG)^2$ and obtain
\bse 
\label{pi3}
\be
|\a\b\r & \lra & |\a\b\r\,,\quad \a\b=0\,,\\[1mm]
|\a\b\r & \lra & e^{i2\vartheta}|\a\b\r\,,\quad \a\b=1\,,
\ee
\ese
where $\vartheta$ is defined in Eq.~(\ref{two8}) and the single-qubit
rotations $S'$ are
\be
\label{pi4}
S'=S_1'\otimes S_2'\,,
\ee
with
\be
\label{pi5}
S_i'=|0\r_i\l 0|e^{-i\vartheta/2}+|1\r_i\l 1|e^{i\vartheta/2}\,,
\ee
where we have dropped a global phase $-(\Theta_{01}+\Theta_{10})/2$.
From Eq.~(\ref{pi3}), the logical
state $|11\r$ is rotated during the new gate sequence by $2\vartheta$, so
to achieve the desired evolution in Eq.~(\ref{two2}) we 
require $2\vartheta=\pi$. This means that in the new sequence
$S'(RG)^2$ each operation $G$ will typically (assuming a linear
dependence in the gate time) be applied for a time half as
long as in the original sequence $SG$.

\subsection{Fidelity}

The calculation of fidelity of a two-qubit phase gate was discussed in
detail in Refs.~\cite{ions3, ions4}. Here we briefly present
the results, in order to
clarify the importance of the $\pi$-pulse method, which will be extended to
a three-ion gate in Sec.~\ref{toffoli}.

We consider a~general initial state
\be
|\Psi_0\r=\sum_{\a,\b=0}^1c_{\a\b}|\a\b\r\,,
\ee
and calculate the fidelity minimised over $c_{\a\b}$, 
i.e. we consider a ``worst-case'' initial state of the qubits.
The fidelity of a two-qubit phase gate corresponding 
to the pulse sequence $SG$ can be expressed then as
\be
{\cal F}
&=&
\bigg\l
\min_{\{c_{\a\b}\}}
\bigg|
\sum_{\a,\b=0}^1|c_{\a\b}|^2e^{i\delta\Theta_{\a\b}}
\bigg|^2
\bigg\r\,,
\ee
where $\delta\Theta_{\a\b}$ is a random departure of the phase $\Theta_{\a\b}$
from its expected value $\l\Theta_{\a\b}\r$ and
$\l\cdot\r$ denotes averaging over all degrees of freedom of the
system. For trapped ions  
it corresponds to averaging over all ion trajectories. 

The fidelity of a two-qubit phase gate corresponding to the sequence 
$S'(RG)^2$ (i.e. phase gate with $\pi$ pulses) can be written as
\be
{\cal F}'=
\bigg\l
\min_{\{c_{\a\b}\}} 
\bigg|
\sum_{\a,\b=0}^1|c_{\a\b}|^2e^{i(\delta\Theta_{\a\b}+\delta\Theta_{\ap\bp})}
\bigg|^2
\bigg\r\,,
\ee
where $\ap\equiv 1-\a$ and $\bp\equiv 1-\b$. Whenever 
$\delta\Theta_{\a\b}\ll 1$ and 
$\delta\Theta_{\a\b}+\delta\Theta_{\ap\bp}\ll\delta\Theta_{\a\b}$
(achieved by applying the $\pi$~pulses),
we obtain a significant improvement in the~fidelity of the~phase
gate using the~$\pi$~pulses, that is $1-{\cal F}'\ll1-{\cal F}$.

\subsection{Implementation on ions in microtraps}
\label{impl}

Consider two ions confined in two separate harmonic potentials
(microtraps), with a logical-state-selective and
time-dependent external force acting on both ions. In the semiclassical
approach (trajectories of the ions are considered to be classical) the
Hamiltonian of the system is
\be
\label{two9}
H(t)=\sum_{\a,\b=0}^1
H_{\a\b}(t)\,|\a\r_1\l\a|\otimes|\b\r_2\l\b|\,,
\ee
where
\be
\label{two10}
H_{\a\b}(t)
&=&
\frac{p_{\a}^2}{2m}+\frac{(p_{\b}^{\p})^2}{2m}\nonumber\\[1mm]
& &
+\frac{1}{2}m\omega^2(x_{\a}+d_0/2)^2+\frac{1}{2}m\omega^2(x_{\b}^{\p}-d_0/2)^2
\nonumber\\[1mm]
& &
+\big(s-x_{\a}-d/2\big)F_{\a}(t) + \big(s^{\p}-x^{\p}_{\b}+d/2\big)F_{\b}'(t)
\nonumber\\[1mm]
& &
+\frac{\ell}{|x_{\b}^{\p}-x_{\a}|}\,,
\ee
where the two bare (empty) microtraps with a~trapping frequency~$\omega$
are separated by a~distance $d_0$, $m$ is the ion mass, 
$x_{\a}(t)$ and $x_{\b}^{\p}(t)$ are coordinates (trajectories) 
of ions~1 and~2 corresponding to their internal states
$|\a=0,1\r_1$ and $|\b=0,1\r_2$, $p_{\a}(t)$ and $p_{\b}^{\p}(t)$ are momenta
of the~ions, and $d$ is the equilibrium distance between the ions ($d>d_0$)
minimizing the total confining potential
(potential of microtraps + Coulomb repulsion).
We denote $\ell=q^2/4\pi\varepsilon_0$, where $q$ is
the ion charge and $\varepsilon_0$ is the~permittivity of vacuum.
Finally, $F_{\a}(t)\equiv\a F(t)$ is a~logical-state-selective ($\a=0,1$)
and time-dependent force 
which displaces the ion only when it is in its logical (internal) 
state~$|\a=1\r$. The~parameters~$s$ and $s^{\p}$ are associated 
with the potential 
(introduced by the displacing forces $F$ and $F'$) which
the~ions experience when they are in their equilibrium positions~($x=\pm d/2$)
in the~microtraps. 

The dynamics of the ion system during the $G$ pulse are governed by the evolution operator
\be
\label{two11}
U=D\exp\left[
-\frac{i}{\hbar}\int_{t_0}^t H(t^{\p})\,dt^{\p}
\right],
\ee
where $D$ denoted the Dyson time-ordering operator, and the Hamiltonian
$H(t)$ is defined in Eq.~(\ref{two9}). The integration is carried out
over a time interval $(t_0,t)$, where we will assume that $t_0<0$, $t>0$,
$|t_0|, |t|\gg\tau$, where $\tau$ describes the duration of the time 
interval on which the pushing force is applied.
This assumption will become clear when we choose the
time profile of the force later on in this section.
Then, the corresponding evolution is
\be
\label{two12}
U|\a\b\r=e^{i\Theta_{\a\b}}|\a\b\r\,,
\ee
where
\be
\label{two13}
\Theta_{\a\b}=-\frac{1}{\hbar}\int_{t_0}^t H_{\a\b}(t^{\p})\,dt^{\p}\,.
\ee
The overall phase in Eq.~(\ref{two8}) can be written
(after a~coordinate transformation $x_{\a}\ra x_{\a}-d/2$ and
$x_{\b}'\ra x_{\b}'+d/2$) as
\be
\label{two14}
\vartheta=\phi_{11}-\phi_{10}-\phi_{01}+\phi_{00}\,,
\ee
where
\be
\label{two15}
\phi_{\a\b}=-\frac{1}{\hbar}\int_{t_0}^t
\frac{\ell}{|d+x_{\b}'-x_{\a}|}\,dt'\,.
\ee
This means that the overall phase of the phase gate is determined only by
the Coulomb interaction between the ions.

Let us define an important parameter
\be
\label{two18}
\epsilon\equiv\frac{q^2}{\pi\varepsilon_0m\omega^2 d^3}\,, 
\ee
which gives the ratio between the Coulomb repulsion energy and the trapping
potential.  It is small ($\epsilon\ll 1$) when the traps (i.e. the ions as
well) are far apart and it can be large ($\epsilon\sim 1$) when we move the
traps close to each other. Here we consider the performance of 
the~phase gate in the regime $\epsilon\ll 1$. 
In Sec.~\ref{general} we extend the treatment to all values of~$\epsilon$.

In the regime $\epsilon\ll 1$ an ion in the internal state $|1\r$ is
displaced from its equilibrium position by
\be
\label{two16}
\bar{x}(t)=\frac{F(t)}{m\omega^2}\,,
\ee
which also corresponds to the displacement of a single ion in a harmonic
trapping potential (Appendix D in Ref.~\cite{ions4}). An ion in the internal
state $|0\r$ is not displaced because the internal-state-selective force
does not act on it. Thus, in the regime $\epsilon\ll 1$ we can write for the
displacement of an ion $\bar{x}_{\a}(t)\equiv\a\bar{x}(t)$, where 
$\a\in\{0,1\}$.
Then, it can be shown \cite{ions3, ions2} that the overall phase 
of the two-qubit phase gate in Eq.~(\ref{two8}) is $\vartheta\approx\theta$
with
\be
\label{two20}
\theta \equiv \sqrt{\frac{\pi}{8}}\,\epsilon\omega\tau\xi^2\,,
\ee
where we assumed a Gaussian time profile of the force, 
$F(t)=F_0 e^{-(t/\tau)^2}$, introduced $\xi=aF_0/\hbar\omega$, and
denoted $a=\sqrt{\hbar/m\omega}$. Then it follows from Eq.~(\ref{two16})
that $\bar{x}(t)=\xi a e^{-(t/\tau)^2}$. This means that the displacement
$\bar{x}$ is measured in units $a$, where $\xi$ is a dimensionless
parameter.
 
The time $\tau$ for which the force needs to be applied is to be
calculated from the condition $\vartheta=\pi$ (i.e. when the phase gate is 
equivalent to the controlled-NOT gate). In practice, 
the internal-state-selective force $F$ will be
produced in a non-resonant laser beam (dipole force) with 
$F \propto I/\Delta$, where $I$ is the laser intensity, and $\Delta$ is
detuning from a driven atomic transition $|1\r\leftrightarrow |\mbox{aux}\r$.
For convenience we will treat the case where 
the logical state $|0\r$ is not coupled to the laser, i.e. ensuring the
state selectivity of the force. The extension to the case where both
logical states experience a non-zero but different force is straightforward.

\subsection{Cancellation of single-qubit rotations}

The $\pi$-pulse method (spin-echo) improves the fidelity of the phase gate
by cancelling some single-qubit rotations which may be imprecise.
Here, we propose two alternative ways to achieve the same effect without
using any $\pi$ pulses. 

The phases $\Theta_{\a\b}$ in Eq.~(\ref{two1}) have a structure~\cite{ions4}
\be
\label{opp1}
\Theta_{\a\b}=
(\a-\b)\theta_1\frac{d}{\xi a}+(\a-\b)^2\theta_2+{\cal O}(a/d)\,,
\ee
where $\theta_1, \theta_2$ are constants of order one
($\theta_1,\theta_2\sim 1$), we have typically $a/d\sim 10^{-3}$, and 
we already know from the last paragraph in the previous section
that $\xi\propto F_0\propto I/\Delta$. 

The second term in Eq.~(\ref{opp1})
is of order one because this term produces the two-particle
phase (\ref{two20}), and the gate time is chosen to ensure 
this phase is $\pi$ or $\pi/2$. With $\xi \sim 1$, 
which it will be seen later is a reasonable value,
the first term is of order $10^3$.

When the laser intensity fluctuates it causes the value of $\xi$ to
fluctuate as well. Let us assume, for example, 1\% laser intensity 
fluctuations.
Then the first term in Eq.~(\ref{opp1}) 
produces an uncertainty of order ten, which is enough
to completely spoil the gate. If this can be suppressed 
then the gate will work
since the second term has a much smaller uncertainty. 
We note that the~combination 
\be
\label{cancel}
\Theta_{01}+\Theta_{10}=2\theta_2+{\cal O}(a^2/d^2)     
\ee
achieves the cancellation while preserving the $\theta_2$ term which is needed 
for the gate. To obtain this cancellation, we can use two pulses 
with opposite signs for
the $\theta_1$ term but the same sign for the $\theta_2$ term. 
The $\pi$-pulse method discussed above
achieves this by swapping the sign of $\alpha-\beta$. An alternative approach
is to swap the sign of $\xi$. Since $\xi \propto 1/\Delta$ this can be done
by using two successive pulses with opposite detuning $\Delta$. This is
our first suggestion. We thus
replace the spin-echo pulse sequence $RGRG$ with the sequence
$G_b G_r$, where $G_b$ ($G_r$) is the $G$ pulse with blue detuning 
(red detuning).

We showed in Ref.~\cite{ions3} that the main contributions to $\theta_1$
are from the Coulomb energy and the light shift (AC Stark shift) produced by
the laser field providing the dipole force. We there derived
the condition for these to be equal and opposite (so called ``sweet spot'').

We next consider how
to cancel them each separately. It is possible to
make the contribution from the light shift zero by a judicious
choice of laser polarization, such that the two internal states
$\ket{0}$ and $\ket{1}$ experience equal light shifts but different
forces \cite{dalibard89, wineland03}. Only the Coulomb contribution
then remains. This can be cancelled by applying the force in
two successive pushes in opposite directions. The linear
$\theta_1$ term in Eq.~(\ref{opp1}) then cancels, while the $\theta_2$ 
term remains.

In all these methods, the 
cancellation takes place as long as the laser intensity is the
same for the two closely-spaced pulses. The gate remains sensitive
to intensity changes between one pulse and its partner in a given
pair, but is much less sensitive to changes in laser intensity between
one pulse-pair and another.

Avoiding the spin-echo method is advantageous since 
the single-qubit $\pi$-pulses it requires may be slow or imprecise. However
there are practical problems in all the methods.
Switching of the detuning
may be technically difficult since $\Delta$ can be of order GHz. 
Switching the force direction is not convenient in a travelling
wave configuration where the force on each ion arises from the transverse
profile of a single focused laser beam, however it
can be done conveniently in a
standing wave configuration by changing the relative phase of the
two travelling waves forming the standing wave \cite{wineland03}.

\section{Decoherence-free subspace}

Several advantages may be obtained by encoding logical qubits not
in single ions but in pairs of ions, in the decoherence-free
subspace (DFS) spanned by
\be
\ket{0}_L\equiv\ket{01}\,,\quad\ket{1}_L\equiv\ket{10}\,.
\ee
Here the subscript $L$ refers to logical states in the subspace,
the kets without subscript refer to physical states of the trapped
ions. This concept has been discussed in the context of trapped ions
in Ref.~\cite{kiel02}. The advantages arise from the
fact that the two logical states evolve in the same way under any
influence which causes the same phase change of $\ket{1}$ relative
to $\ket{0}$ in both ions, therefore the logical information is
completely protected from such influences. An example is a change
in the magnetic field on both ions together, or a phase drift of
the lasers used to implement the operations.

\begin{figure}[htb]
\includegraphics[width=8.5cm]{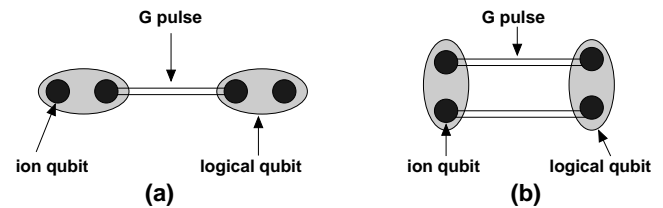}
\caption{Two different ways of encoding logical qubits in pairs of
ions qubits in the decoherence-free subspace (DFS).
(a)~Two pairs of ion qubits are arranged along a line to form 
a~pair of logical qubits. We apply the~$G$ pulse only between 
the two central ions. (b) A more symmetric way of encoding logical qubits in
the DFS. We apply the $G$ pulse between the two pairs of ion qubits.}
\label{dfs_fig}
\end{figure}

We will next present two methods to implement the controlled-phase gate $\C Z$
in the DFS. The required evolution is
\be
\label{dfs}
\begin{array}{lcr}
\ket{01}\ket{01} & \rightarrow & \ket{01}\ket{01}\,,\\[1mm]
\ket{01}\ket{10} & \rightarrow & \ket{01}\ket{10}\,,\\[1mm]
\ket{10}\ket{01} & \rightarrow & \ket{10}\ket{01}\,,\\[1mm]
\ket{10}\ket{10} & \rightarrow & -\ket{10}\ket{10}\,.
\end{array}
\ee

First, consider two pairs of ions arranged along a line as
in FIG.~\ref{dfs_fig}a. They may be all in separate microtraps, or
with one or more in the same trap.
All we need to do to achieve the desired gate is apply the pushing
method to the central two ions. The presence of the other ions has
a small influence on the distance through which the pushed ions
are displaced, and their Coulomb interactions contribute to the
total phase. However, this simply changes by a factor of order one
the time required to achieve the desired phase $\pi$, but it does not
introduce sensitivity to thermal motion.
We thus obtain the evolution
\be
\label{dfs1}
\begin{array}{lcr}
\ket{01}\ket{01} & \rightarrow & \ket{01}\ket{01}\,,\\[1mm]
\ket{01}\ket{10} & \rightarrow & -\ket{01}\ket{10}\,,\\[1mm]
\ket{10}\ket{01} & \rightarrow & \ket{10}\ket{01}\,,\\[1mm]
\ket{10}\ket{10} & \rightarrow & \ket{10}\ket{10}\,.
\end{array}
\ee
To obtain the phase gate in Eq.~(\ref{dfs}),
the Pauli $\sigma_Z$ operation is then applied 
to the second logical qubit.

It may be objected that the operation just described lacks symmetry
in the way the controlling apparatus interacts with the ions in a given
pair, and therefore it is against the spirit of the
DFS concept. However, the joint state of the
ions remains at all times in the DFS, and the mechanism of
the pushing force (such as a far off-resonant dipole force)
does not rely on the need for the very precise frequency
matching which the DFS is designed to avoid. Therefore the need
for very precise frequency matching (or magnetic field control) 
is still avoided.

An alternative and more symmetric way to implement the gate in
the DFS is shown in FIG.~\ref{dfs_fig}b~\footnote{The original idea comes
from T. Calarco.}.
By pushing the ions as shown,
before any single-qubit rotations are added
the evolution is
\be
\label{df2}
\begin{array}{lcl}
\ket{01}\ket{01} & \rightarrow &
e^{i(\Theta_{00} + \Theta_{11})}\ket{01}\ket{01}\,,\\[1mm]
\ket{01}\ket{10} & \rightarrow &
e^{i(\Theta_{01} + \Theta_{10})}\ket{01}\ket{10}\,,\\[1mm]
\ket{10}\ket{01} & \rightarrow &
e^{i(\Theta_{10} + \Theta_{01})}\ket{10}\ket{01}\,,\\[1mm]
\ket{10}\ket{10} & \rightarrow &
e^{i(\Theta_{11} + \Theta_{00})}\ket{10}\ket{10}\,,
\end{array}
\ee
where $\Theta_{\alpha\beta}$ are the same phases
as in Eq.~(\ref{two1}). Recalling Eq.~(\ref{two8}),
the overall phase for the evolution of the logical qubits is
\be
\vartheta
&=&
(\Theta_{11} + \Theta_{00})
-2(\Theta_{10} + \Theta_{01})
+(\Theta_{00} + \Theta_{11})\nonumber\\[1mm]
&=&
2(\Theta_{11} - \Theta_{10} - \Theta_{01} + \Theta_{00})\,.
\ee
Therefore the gate is obtained in a time half that required for
single ions in the same conditions.

The method of FIG.~\ref{dfs_fig}b may be more
difficult to realise in practice, owing to the two-dimensional
rather than one-dimensional configuration of the ions. However, it
has the advantage that the cancellation of unwanted phases
in Eq.~(\ref{opp1}) takes place automatically
without any need for $\pi$~pulses or switched detunings. This also
means the ``sweet spot'' method discussed in Ref.~\cite{ions3}
is not needed.

\section{Non-neighbour ions and Toffoli gate}
\label{toffoli}

\subsection{Principle}

We now consider a system of three qubits. First we present
an analysis of the general requirements along similar lines to that
used for two qubits in Sec.~\ref{two}. Suppose
a~system of three interacting qubits evolves as
\be
\label{toff1}
|\a\b\g\r\stackrel{\tilde{G}}{\ \lra\ }e^{i\Theta_{\a\b\g}}|\a\b\g\r\,,
\ee
where $|\a\b\g\r\equiv |\a\r_1\otimes |\b\r_2\otimes |c\r_3$ denotes a joint
logical state of three qubits with $\a,\b,\g\in\{0,1\}$. 
Our aim is now to realize a three-qubit phase gate
\bse
\label{toff2}
\be
|\a\b\g\r &\lra& |\a\b\g\r\,,\quad \a\b\g=0\,,\\
|\a\b\g\r &\lra& e^{i\tilde{\vartheta}}|\a\b\g\r\,,\quad \a\b\g=1\,,
\ee
\ese
that is the gate which produces an overall phase $\tilde{\vartheta}$
if and only if all three qubits are in the logical state $|1\r$.
The three-qubit phase gate can be obtained when we apply single-qubit
rotations after the operation $\tilde{G}$, that is
\be
\label{toff3}
\tilde{S}=\tilde{S}_1\otimes\tilde{S}_2\otimes\tilde{S}_3\,,
\ee
where
\bse
\label{toff4}
\be
\tilde{S}_1&=&
|0\r_1\l 0|\,e^{iA_0}+
|1\r_1\l 1|\,e^{iA_1}\,,\\[1mm]
\tilde{S}_2&=&
|0\r_2\l 0|\,e^{iB_0}+
|1\r_2\l 1|\,e^{iB_1}\,,\\[1mm]
\tilde{S}_3&=&
|0\r_3\l 0|\,e^{iC_0}+
|1\r_3\l 1|\,e^{iC_1}\,.
\ee
\ese
The sequence $\tilde{S}\tilde{G}$
will correspond to the
three-qubit gate if the following equalities are satisfied
\bse
\label{toff5}
\be
A_0+B_0+C_0+\Theta_{000} &=& 0\,,\\
A_1+B_0+C_0+\Theta_{100} &=& 0\,,\\
A_0+B_1+C_0+\Theta_{010} &=& 0\,,\\
A_0+B_0+C_1+\Theta_{001} &=& 0\,,\\
A_1+B_1+C_0+\Theta_{110} &=& 0\,,\\
A_1+B_0+C_1+\Theta_{101} &=& 0\,,\\
A_0+B_1+C_1+\Theta_{011} &=& 0\,,
\ee
\ese
where the overall phase reads
\be
\label{toff6}
\tilde{\vartheta}=A_1+B_1+C_1+\Theta_{111}\,.
\ee
In Eqs.~(\ref{toff5}) we have seven equations in
six independent parameters ($A_0,\dots,C_1$), but the equations
are not all linearly independent. By combining them we obtain
\bse
\label{toff7}
\be
\Theta_{000}+\Theta_{110} &=& \Theta_{100} + \Theta_{010}\,,\\
\Theta_{000}+\Theta_{101} &=& \Theta_{100} + \Theta_{001}\,,\\
\Theta_{000}+\Theta_{011} &=& \Theta_{010} + \Theta_{001}\,,
\ee
\ese
These are conditions on the evolution (\ref{toff1}) which will
have to be satisfied if the Toffoli gate (the three-qubit phase gate, 
to be precise) is to be realised.

When we apply Eqs.~(\ref{toff7}),
then Eqs.~(\ref{toff5}e)--(\ref{toff5}g) can be expressed as
linear combinations of Eqs.~(\ref{toff5}a)--(\ref{toff5}d). This means that
the original system~(\ref{toff5}) is now reduced to four independent
equations and six independent parameters. Therefore, we can choose
two parameters and calculate the rest. Let us pick
$A_0=B_0=-\Theta_{000}/3$. We then get $C_0=-\Theta_{000}/3$ and
\bse
\label{toff8}
\be
A_1 &=& -\Theta_{100}+2\Theta_{000}/3\,,\\
B_1 &=& -\Theta_{010}+2\Theta_{000}/3\,,\\
C_1 &=& -\Theta_{001}+2\Theta_{000}/3\,,
\ee
\ese
where the overall phase is
\be
\label{toff9}
\tilde{\vartheta}=
\Theta_{111}-\Theta_{100}-\Theta_{010}-\Theta_{001}+2\Theta_{000}\,.
\ee

We can also introduce $\pi$-pulses in order to cancel unwanted terms.
We replace
the sequence $\tilde{S}\tilde{G}$ by $\tilde{S}'(\tilde{R}\tilde{G})^2$,
where $\tilde{R}=R_1\otimes R_2\otimes R_3$
denotes a $\pi$ pulse applied on all three qubits. An analysis along
similar lines to the above implies that the three-qubit phase is now given by
\be
\label{toff_pi2}
\tilde{\vartheta}'&=&
3(\Theta_{000}+\Theta_{111})\nonumber\\[1mm]
&&
-(\Theta_{100}+\Theta_{011}+\Theta_{010}+\Theta_{101}
+\Theta_{001}+\Theta_{110})\,.\nonumber\\
\ee
The single-qubit rotations 
$\tilde{S}'=\tilde{S}'_1\otimes\tilde{S}'_2\otimes\tilde{S}'_3$
take the same form as in Eqs.~(\ref{toff4}), except that now we have 
$A'_0=B'_0=C'_0=-(\Theta_{000}+\Theta_{111})/3$, and
\bse
\label{toff_pi3}
\be
A'_1 &=& -(\Theta_{100}+\Theta_{011})+2(\Theta_{000}+\Theta_{111})/3\\[1mm]
B'_1 &=& -(\Theta_{010}+\Theta_{101})+2(\Theta_{000}+\Theta_{111})/3\\[1mm]
C'_1 &=& -(\Theta_{001}+\Theta_{110})+2(\Theta_{000}+\Theta_{111})/3\,.
\ee
\ese

There is again a set of conditions to be satisfied 
if the three-qubit phase gate with $\pi$ pulses is to be realized. They are
\bse
\label{toff_pi4}
\be
\Theta_{110} + \Theta_{001} + \Theta_{000} + \Theta_{111} 
&=&
\Theta_{100} + \Theta_{011} + \Theta_{010} + \Theta_{101}\,,\nonumber\\
\\
\Theta_{101} + \Theta_{010} + \Theta_{000} + \Theta_{111} 
&=&
\Theta_{100} + \Theta_{011} + \Theta_{001} + \Theta_{110}\,,\nonumber\\
\\
\Theta_{011} + \Theta_{100} + \Theta_{000} + \Theta_{111} 
&=&
\Theta_{010} + \Theta_{101} + \Theta_{001} + \Theta_{110}\nonumber\\
\ee
\ese

Thus, the general concept of the three-qubit phase gate 
(which is equivalent to the Toffoli gate) seems to work in both cases.
The question is now whether we are able to satisfy
Eqs.~(\ref{toff7}) or (\ref{toff_pi4}), respectively, 
on the system of ions in microtraps. We discuss this
issue in the following section.

\subsection{Implementation on ions in microtraps}

The Hamiltonian for the system
of three ions in three separate microtraps is
\be
\label{toff10}
H(t)=\!\!\sum_{\a,\b,\g=0}^1
H_{\a\b\g}(t)\,|\a\r_1\l\a|\otimes|\b\r_2\l\b|\otimes|\g\r_3\l\g|,
\ee
with
\be
\label{toff11}
\leftline{$H_{\a\b\g}(t)=$}
\nonumber\\[1mm]
\sum_{j=1}^3\left\{
\frac{p_{\a_j}^2}{2m}
+\frac{1}{2}m\omega^2
\left[(x_{\a_j}-\bar{x}_{\a_j})^2-\bar{x}_{\a_j}^2+2\bar{x}_{\a_j}s_j\right]
\right\}
\nonumber\\
+\frac{\ell}{|d+x_{\b}^{\p}-x_{\a}|}
+\frac{\ell}{|2d+x_{\g}''-x_{\a}|}
+\frac{\ell}{|d+x_{\g}''-x_{\b}'|}\,,
\nonumber\\
\ee
where we introduced a notation $\a_1\equiv\a$, $\a_2\equiv\b$,
$\a_3\equiv\g$ and $x_{\a_1}\equiv x_{\a}$, $x_{\a_2}\equiv x_{\b}'$,
$x_{\a_3}\equiv x_{\g}''$, and we applied a coordinate transformation
$x_{\a}\to x_{\a}-d$, $x_{\b}'\to x_{\b}'$ and $x_{\g}''\to x_{\g}''+d$
assuming that the ions are at sites $x=-d$, $x=0$ and $x=d$ with
$\epsilon\ll 1$. Then the three-particle dynamical phases 
$\Theta_{\a\b\g}$ read
\be
\label{toff12}
\Theta_{\a\b\g}=-\frac{1}{\hbar}\int_{t_0}^t H_{\a\b\g}(t')\,dt'\,.
\ee
The structure of the Hamiltonian in Eq.~(\ref{toff11}) allows us to write
for the overall phase
\be
\label{toff13}
\tilde{\vartheta}=\phi_{111}-\phi_{100}-\phi_{010}-\phi_{001}+2\phi_{000}\,,
\ee
where
\be
\label{toff14}
\phi_{\a\b\g}&=&   -\frac{1}{\hbar}\int_{t_0}^t
\left\{
\frac{\ell}{|d+x_{\b}^{\p}-x_{\a}|}
+\frac{\ell}{|2d+x_{\g}''-x_{\a}|}
\right.\nonumber\\[1mm]
& &
\left.
+\frac{\ell}{|d+x_{\g}''-x_{\b}'|}
\right\}
dt'.
\ee
It follows from Eq.~(\ref{toff13}) that the overall phase is determined
again only by the interaction phases $\phi_{\a\b\g}$ rather than by the
total phases $\Theta_{\a\b\g}$. However, in contrast to Eq.~(\ref{two15})
this time we have to consider three pairs of mutual Coulomb
interactions.

We need to verify whether Eqs.~(\ref{toff7}) can be satisfied for
the system of ions in microtraps. In the adiabatic approximation
($\omega\tau\gg 1$) and for $\epsilon\ll 1$ we can
write the trajectories of the ions in the form
\bse
\label{toff15}
\be
x_{\a} &\approx& \bar{x}_{\a}+\delta\,,\\
x_{\b}' &\approx& \bar{x}_{\b}'+\delta'\,,\\
x_{\g}'' &\approx& \bar{x}_{\g}''+\delta''\,,
\ee
\ese
where $\bar{x}_{\a}$ denotes the state-selective displacement of the ion 
(Eq.~(\ref{two16})), and $\delta$ corresponds to thermal
oscillations of the ion in the microtrap.
The adiabatic regime means that the state-selective pushing force $F$ 
is applied on 
a time scale much longer than an oscillation period of the ions in microtraps.
We assume first of all that state-dependent
forces of the same magnitude and the same pulse time $\tau$ 
are applied to all three ions.
Then, the analysis of Eq.~(\ref{toff14}), using the methods of
\cite{ions3,ions4}, shows that the three-particle phases take the form
\be
\label{toff16}
\phi_{\a\b\g}\approx
-\frac{\theta}{2}
\left[
(\a-\b)^2+\frac{1}{8}(\a-\g)^2+(\b-\g)^2
\right]\,,
\ee
where $\theta$ is defined in Eq.~(\ref{two20}), $\a,\b,\g\in\{0,1\}$, and we
dropped terms of order ${\cal O}(a^2/d^2)$ because we have typically
$a/d\sim 10^{-2}-10^{-3}$.

When we apply the~gate pulse $\tilde{G}$ on three ions 
for a time $\tau$, followed by the~single-qubit rotations $\tilde{S}$, 
we obtain
\be
\label{toff17}
\begin{array}{llllll}
|000\r & \stackrel{\tilde{S}\tilde{G}}{\longrightarrow} & |000\r\,,
&
\qquad
|110\r & \longrightarrow & e^{i\theta}|110\r\,,\\[1mm]
|100\r & \longrightarrow & |100\r\,,
&
\qquad
|101\r & \longrightarrow & e^{i\theta/8}|101\r\,,\\[1mm]
|010\r & \longrightarrow & |010\r\,,
&
\qquad
|011\r & \longrightarrow & e^{i\theta}|011\r\,,\\[1mm]
|001\r & \longrightarrow & |001\r\,,
&
\qquad
|111\r & \longrightarrow & e^{i 17\theta/8}|111\r\,,
\end{array}
\ee
which corresponds to 
$^C\!P_{12}(\theta) \, ^C\!P_{23}(\theta) \, ^C\!P_{13}(\theta/8)$,
where $^C\!P_{ij}(\theta)$ 
is the phase gate in Eq.~(\ref{first}) between ions~$i$ and~$j$. 
This means that the single
pulse $\tilde{G}$ on all three ions simultaneously 
(followed by single-qubit rotations $\tilde{S}$)
produces three two-qubit controlled-phase gates. 
The two-qubit phase of the gate between the outermost
pair of ions is $\theta/8$, rather than $\theta$, because the separation
of those ions is $2d$ and the two-qubit gate phase is proportional 
to the inverse cube of the separation.

\begin{figure}[htb]
\includegraphics[width=8cm]{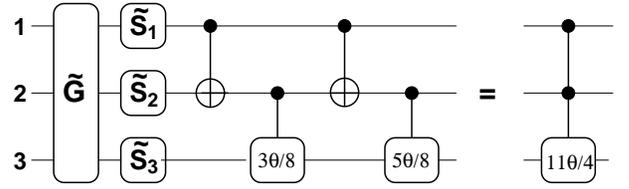}
\caption{Three-qubit phase gate. A $\tilde{S}\tilde{G}$ pulse sequence 
followed by a quantum-gate network produces a three-qubit phase gate 
when $\theta=4\pi$.
The network corresponds to the transformation in Eq.~(\ref{toff18}).}
\label{toff_fig1}
\end{figure}

To obtain the simple three-qubit phase gate $\cc Z$ from Eq.~(\ref{toff17})
we apply a network as in FIG.~\ref{toff_fig1}. 
A sequence containing
two controlled-NOT and two controlled-phase gates on adjacent ions
causes the overall evolution to be
\be
\label{toff18}
\begin{array}{llllll}
|000\r & \longrightarrow & |000\r\,, & \qquad
|110\r & \longrightarrow & e^{i\theta}|110\r\,,\\[1mm]
|100\r & \longrightarrow & |100\r\,, & \qquad
|101\r & \longrightarrow & e^{i\theta/2}|101\r\,,\\[1mm]
|010\r & \longrightarrow & |010\r\,, & \qquad
|011\r & \longrightarrow & e^{i 2 \theta}|011\r\,,\\[1mm]
|001\r & \longrightarrow & |001\r\,, & \qquad
|111\r & \longrightarrow & e^{i11\theta/4}|111\r\,,
\end{array}
\ee
which is a three-qubit phase gate $^{CC}Z$ for $\theta=4\pi$ 
(or an integer multiple of $4 \pi$). The controlled-NOT gates can be obtained
from the two-qubit phase gate plus Hadamard rotations (or $\pi/2$ pulses).

Note that when $\theta=4 \pi$,
the operation (\ref{toff17}) is $^C\!P_{13}(\pi/2)$. 
In other words, we have simply a two-qubit phase gate between 
the outermost pair of ions. In this case
the force on the central ion is redundant, i.e. 
the same result would be obtained with
laser beams pushing only the two outermost ions.
This also shows that the network of FIG.~\ref{toff_fig1} is 
a standard decomposition of $\cc Z$ as presented in Ref.~\cite{barenco95}.

\begin{figure}[htb]
\includegraphics[width=8cm]{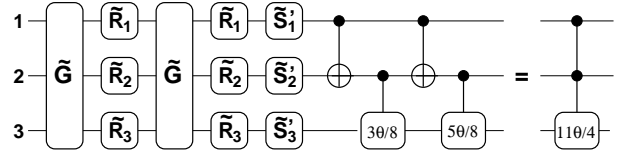}
\caption{Three-qubit phase gate with $\pi$ pulses. 
A $\tilde{S}'(\tilde{R}\tilde{G})^2$ pulse sequence,
followed the same network as in FIG.~\ref{toff_fig1}, 
produces a three-qubit phase gate with $\pi$ pulses, 
i.e. a gate with a better 
performance in terms of fidelity. Note that each $\tilde{G}$ gate 
in this network is
obtained by applying the pushing force for half as long as for $\tilde{G}$ 
in FIG.~\ref{toff_fig1}.}
\label{toff_fig2}
\end{figure}

The implementation of the phase gate including a spin-echo by
$\pi$ pulses is also completed by the same network 
(FIG.~\ref{toff_fig2}).

We note that in principle, two-qubit gates between pairs of ions 
even further apart
could also be implemented simply be pushing the relevant pair of ions, 
without the need first to make them adjacent.

\subsection{Speed of the gate}


Next we estimate the speed of three-qubit phase gates with ions
in microscopic traps. We will consider single-qubit gates 
($\pi$~pulses, rotations) to be fast compared to
multi-qubit gates. This is typically the case for trapped ion
qubits. Therefore we only need to add together the times for 
two-qubit and three-qubit logic gates on the ions. 
We assume that these are all produced
by the state-selective pushing method under discussion.

There are two different assumptions about the speed of the operations
which are both physically sensible. First, we may assume that all operations
involve pushing forces of the same magnitude. In this case the time required
to produce the two-qubit phase gate between the outermost ions is eight 
times longer than
the time to produce the same gate between adjacent ions. On the other hand,
we might assume that the forces are always adjusted in magnitude so that
the gate duration is limited by the adiabatic criterion $\omega\tau\gg 1$. 
In this case the gate between outermost ions has the same duration as
the same gate between adjacent ions, but requires larger forces. 

With the former assumption (forces of given magnitude), the total time
required for the whole network, including $\tilde{G}$
and the two-qubit gates, is 
\be
\label{toff19}
\tau_{\rm total}\approx
4 \tau +\tau+\frac{3\tau}{2}+\tau+\frac{\tau}{2}
= 8 \tau \,,
\ee
where $\tau$ is the time required for a $^C\! Z$ gate (i.e. $\theta=\pi$)
between adjacent ions. The duration of the $\tilde{G}$ pulse is $4\tau$
because we need $\theta=4\pi$. The duration of the final gate 
is $\tau/2$ since $\exp(i\phi)=\exp[i(\phi+2\pi)]$.

With the other assumption (all gates of duration $\tau$), the total time is
$5 \tau$. 

The same conclusions apply to the three-qubit phase gate with $\pi$~pulses
in FIG.~\ref{toff_fig2}, except that now each of the two $\tilde{G}$ 
pulses is applied for a time $2\tau$ (instead of $4\tau$), 
and the fidelity of the gate is higher.

\section{Two-qubit Phase gate in general}
\label{general}

In Ref.~\cite{ions3} we discussed the dynamics of a two-qubit
phase gate with ions in microtraps in the regime $\epsilon\ll 1$, 
that is when
the two microtraps were sufficiently far apart. In what follows we will
re-examine the dynamics of the ion system outside this regime, i.e. 
$\epsilon\sim 1$. We will use a more general approach compared to the one in
Ref.~\cite{ions3}. The aim is to understand the phase gate for all values
of $\epsilon$, and especially for the case of two ions in the same trap, where
$\epsilon=2$.

To this end it is useful to write the Hamiltonian in Eq.~(\ref{two10}) in
a new coordinate system
\bse
\be
\label{lin1}
R_{\a\b}&=&(x_{\a}+x_{\b}')/2\,,\\[1mm]
r_{\a\b}&=&x_{\b}'-x_{\a}-d\,,
\ee
\ese
where $R_{\a\b}$ is a centre-of-mass coordinate of the ion system,
$r_{\a\b}$ is a relative excursion of the ions from their equilibrium 
positions, and $\a,\b\in\{0,1\}$ correspond to logical states
of ion qubits $|0\r, |1\r$. Then, the Hamiltonian splits 
into a centre-of-mass and a relative-motion part
\be
\label{lin2}
H_{\a\b}&=&H_{\a\b}^R+H_{\a\b}^r\nonumber\\
&=&
\frac{P^2_{\a\b}}{2M}+V_{\a\b}^R
+\frac{p^2_{\a\b}}{2\mu}+V_{\a\b}^r\,,
\ee
where we have introduced 
$P_{\a\b}=M\dot{R}_{\a\b}$, $p_{\a\b}=\mu\dot{r}_{\a\b}$, $M=2m$, and
$\mu=m/2$. The potential-energy terms of the two ions in a joint internal
state $|\a\b\r$ are
\bse
\label{lin3}
\be
V_{\a\b}^R
&=&
\frac{1}{2}M\omega^2R^2_{\a\b}
-R_{\a\b}(F_{\a}+F'_{\b})
+(sF_{\a}+s'F'_{\b})\,,
\nonumber\\ \\
V_{\a\b}^r
&=&
\frac{1}{2}\mu\omega^2(r_{\a\b}+\Delta d)^2
-\frac{1}{2}r_{\a\b}(F'_{\b}-F_{\a})+\frac{\ell}{d+r_{\a\b}}\,,
\nonumber\\
\ee
\ese
where we denoted $\Delta d=d-d_0$ (i.e. the difference between
the separation of ions loaded in the traps and the separation of
the centres of bare traps), and we assumed
$|d + r_{\a\b}| = d + r_{\a\b}$ since
$d \gg |r_{\a\b}| > 0$.

The ion separation $d$ can be calculated from the equilibrium 
when no force is applied, that is
\be
\label{lin5}
\left.
\frac{\partial V}{\partial r}
\right|_{r=0}
=
\mu\omega^2\Delta d -\frac{\ell}{d^2}=0\,,
\ee
giving
\be
\label{lin6}
\Delta d=\frac{4d_0}{3}\sinh^2
\left\{\frac{1}{6}\ln\left[\eta+1+\sqrt{\eta(\eta+2)}\right]\right\},
\ee
where $\eta=\ell/[m\omega^2(d_0/3)^3]$. 
Using Eq.~(\ref{lin6}) and substituting $d=d_0+\Delta d$ into 
the expression of $\epsilon$ in Eq.~(\ref{two18}), 
we find that $\epsilon$ is 
a~function of a single variable $\eta$.
The function is monotonic and there are two
limiting cases. (i) When $\eta\to 0$ (which corresponds to $d_0\to\infty$, i.e.
microtraps are far apart), we get 
$\epsilon=q^2/(\pi\varepsilon_0m\omega^2d_0^3)$. This means that
$\epsilon\ll 1$, and $d=d_0$. (ii) On the other hand, when $\eta\to\infty$
(corresponding to $d_0\to 0$, i.e. the two separate microtraps 
overlap completely
and we end up with two ions in a single linear trap), we get $\epsilon=2$,
which is also the maximum value of $\epsilon$.

Substituting $\epsilon=2$ into Eq.~(\ref{two18}) gives
\be
\label{lin7}
d=
\left(
\frac{q^2}{2\pi\varepsilon_0m\omega^2}
\right)^{1/3}\,,
\ee
which is the standard expression of the separation of two ions in the same
harmonic trap. We refer to this case as the case of a ``linear trap'', since
typically one chooses a linear geometry in the Paul trap in order to
avoid micromotion when more than one ion is trapped.
Note that whereas in separate microtraps we could adjust the trapping
frequency and ion separation independently, in a linear trap 
(i.e. two ions in the same trap) they are mutually dependent.

\subsection{Anharmonicity}
\label{anharm}

In the discussion about ions in two separate microtraps
in Sec.~\ref{two}, the results were
valid in the regime $\epsilon\ll 1$. Next we examine the meaning
of this condition, and show how it can be substantially relaxed.
This is necessary in order to allow the gate to be implemented
for two ions in the same trap, since then $\epsilon=2$.

We can expand the Coulomb term in Eq.~(\ref{lin3}b) in a Taylor series
owing to $|r_{\a\b}|\ll d$, giving
\be
\label{lin8}
V_{\a\b}^r
&\approx&
\frac{1}{2}\mu\tilde{\omega}^2r_{\a\b}^2
\left[
1-\frac{\epsilon}{\epsilon+1}
\frac{r_{\a\b}}{d}
\left(
1-\frac{r_{\a\b}}{d}
\right)
\right]\nonumber\\[1mm]
&&
-\frac{1}{2}r_{\a\b}(F'_{\b}-F_{\a})\,,
\ee
where we denoted $\tilde{\omega}=\omega\sqrt{1+\epsilon}$,
used the definition~of~$\epsilonÂ$ in Eq.~(\ref{two18}), 
applied Eq.~(\ref{lin5}), and
dropped a constant term $V_0=\mu\omega^2(\Delta d)^2/2+\ell/d$.
This means that in the regime $\epsilon\sim 1$ the effective trapping
potential becomes anharmonic.  
We need to find out how the anharmonicity affects the centre-of-mass and the
relative motion of the ions.

\subsection{Dynamics}    
\label{dynamic}

Eqs.~(\ref{lin3}) and (\ref{lin8}) lead to the equations of motion
\bse
\label{lin9}
\be
\ddot{R}_{\a\b}+\omega^2R_{\a\b}&=&F_{\a\b}/M\,,\\[1mm]
\ddot{r}_{\a\b}+\tilde{\omega}^2r_{\a\b}
\left(
1-\frac{\epsilon}{\epsilon+1}\frac{3r_{\a\b}}{2d}
\right)
&=&f_{\a\b}/\mu\,,
\ee
\ese
where $F_{\a\b}=F_{\a}+F'_{\b}$, 
$f_{\a\b}=(F'_{\b}-F_{\a})/2$
with $\a,\b\in\{0,1\}$.
In the following we assume for convenience that
the state-selective-force $F_{\a}$ acts only when the ion is 
in its internal state $|\a=1\r$, so that
$F_{\a\b} = (\a+\b)F$, $f_{\a\b} = (\b-\a)F/2$, 
where $F \equiv F_1=F'_1$. Using Appendix~D 
in Ref.~\cite{ions4} the solution of Eq.~(\ref{lin9}a) in the adiabatic
regime ($\omega\tau\gg 1$) is
\be
\label{lin10}
R_{\a\b}(t)\approx
\bar{R}_{\a\b}(t)+\Delta (t)\,,
\ee
with
\bse
\label{lin11}
\be
\bar{R}_{\a\b}(t)
&=&
\frac{F_{\a\b}(t)}{M\omega^2}=(\a+\b)\frac{\bar{x}(t)}{2}\,,
\\[1mm]
\Delta(t)
&=&
\sqrt{\frac{2E_R}{M\omega^2}}\cos(\omega t+\psi_R)\,,
\ee
\ese
where
we used Eq.~(\ref{two16}), and $E_R$ ($\psi_R$) is the oscillation
energy (initial motional phase) of the centre-of-mass motion of the ions.

There is no analytical solution of Eq.~(\ref{lin9}b). However, owing to the
fact that $|r_{\a\b}|\ll d$ the anharmonicity is small even for
$\epsilon\sim 1$. Therefore, we can estimate the solution in the adiabatic
approximation ($\tilde{\omega}\tau\gg 1$) as
\be
\label{lin12}
r_{\a\b}(t)\approx
\bar{r}_{\a\b}(t)+\delta_{\a\b}(t)\,,
\ee
with
\bse
\label{lin13}
\be
\bar{r}_{\a\b}(t)
&=&
\frac{f_{\a\b}(t)}{\mu\tilde{\omega}^2}
\left[
1+\frac{3\epsilon}{2(\epsilon+1)}
\frac{f_{\a\b}(t)}{\mu\tilde{\omega}^2d}+\dots
\right]\nonumber\\[1mm]
&\approx&
(\b-\a)\frac{\bar{x}(t)}{\epsilon+1}\,,\\[1mm]
\delta_{\a\b}(t)
&=&
\sqrt{\frac{2E_r}{\mu\tilde{\omega}^2}}\cos(\tilde{\omega}t+\psi_r)
+\zeta_{\a\b}(t)\,,
\ee
\ese
where $\bar{r}_{\a\b}$ arises from the condition 
$\partial V_{\a\b}/\partial r_{\a\b}=0$,
$E_r$ and $\psi_r$ refer to the oscillation
energy and an initial motional phase of the relative motion of the ions,
and $\zeta_{\a\b}$ is to be discovered.

When only one ion is in the state $|1\r$, that ion is displaced
by the quantity $\bar{x}(t)$ imposed by the force in Eq.~(\ref{two16}).
It follows from Eq.~(\ref{lin13}a) that when $\epsilon\ll 1$ this is also 
the change in the relative separation, i.e.
the other ion hardly moves besides random thermal oscillations. However, in
the regime $\epsilon\sim 1$ the ions are close enough to be strongly 
coupled. Then the other ion moves as well and the net result is that
the relative displacement is reduced by a factor $1+\epsilon$.

The main consequence of anharmonicity is the
extra contribution $\zeta_{\a\b}(t)$ to the relative motion of the ions
in Eq.~(\ref{lin13}b). We no longer have the special property of a harmonic
potential that the average excursion under free oscillatory motion is zero
(when averaged over a large or integral number of oscillations). Instead
the average is $\left< \zeta_{\a\b} \right>$.
We do not have an analytical solution and so we must estimate
$\zeta_{\a\b}(t)$. As an estimate we use the average of the two
turning points of the motion (at given energy) in the anharmonic well,
i.e. $\zeta_{\a\b}=(r'_{\a\b}+r''_{\a\b})/2$, 
where $r'_{\a\b}$ and $r''_{\a\b}$ are the two roots (closest to the
equilibrium point $\bar{r}_{\a\b}$) of the equation
\be
\label{lin14}
V_{\a\b}^r(r_{\a\b})=E_r+V_{\a\b}^r(\bar{r}_{\a\b})\,,
\ee
where $V_{\a\b}^r$ is the anharmonic potential defined in Eq.~(\ref{lin8}),
and $E_r$ is thermal energy of the relative motion of the ions in
Eq.~(\ref{lin13}b). We expect this estimate to be good up to a numerical
factor of order one. By obtaining the two roots and expanding $\zeta_{\a\b}$
as a Taylor series in $E_r$, we find
\be
\label{lin15}
\zeta_{\a\b}(t)\approx
\frac{\epsilon}{\epsilon+1}\frac{E_r}{\mu\tilde{\omega}^2d}
\left[
1+\frac{6\epsilon}{\epsilon+1}\frac{f_{\a\b}(t)}{\mu\tilde{\omega}^2d}
\right]\,.
\ee
Note that the influence of $\zeta_{\a\b}$ on the phases appearing in the
quantum logic gate arises only from that part of $\zeta_{\a\b}$ which
depends on $\alpha$ and $\beta$, the rest will merely add a global phase.
A numerical solution of the equations of motion 
confirmed the main points of this estimate. It was found that
$\l\zeta_{\a\b}\r$ typically underestimates $\l\delta_{\a\b}\r$
by a factor approximately six.

In order to calculate the dynamical phases in Eq.~(\ref{two13})
it is crucial to include both the centre-of-mass motion
and the relative motion, in particular because the two-qubit phase is
found to arise from a difference between the two contributions,
associated with the fact that $\tilde{\omega} \ne \omega$.
By substituting
Eqs.~(\ref{lin10}) and~(\ref{lin12}) into the Hamiltonian (\ref{lin2}) we
determine the dynamic phases in Eq.~(\ref{two13}) and hence obtain the
overall phase in Eq.~(\ref{two8}). The result is
\be
\label{lin16}
\vartheta=\frac{\theta}{\epsilon+1}
\left[
1-\frac{1}{(\omega\tau)^2}\frac{2+\epsilon}{1+\epsilon}
+{\cal O}(a^2/d^2)
\right]\,.
\ee
where $\theta$ is given by Eq. (\ref{two20}).

In the limit when $\epsilon\ll 1$ and $\omega\tau\gg 1$ we obtain
$\vartheta\approx\theta$ as before. 
The main effect of  bringing the traps together is to introduce
the factor $1/(\epsilon+1)$.
The (small) term in $1/(\omega \tau)^2$ in Eq.~(\ref{lin16}) is a surprise since
it survives even when $\epsilon \rightarrow 0$, and therefore it should
have been present in the earlier calculations. It arises from the fact 
that the kinetic energy of the vibrational
motion depends on where the ion is in the harmonic well, and we now allow
for the fact that $\tilde{\omega}\ne\omega$ in all kinetic
energy terms whereas previously we did not.

\subsection{Phase gate in a linear trap}

Let $\vartheta_L$ be the overall two-qubit phase from 
Eq.~(\ref{lin16}) in the case of two ions in the same
linear ion trap ($\epsilon=2$). Then
\be
\label{lin17}
\vartheta_L
\approx\theta/3
=\frac{2}{3}\sqrt{\frac{\pi}{8}}\omega\tau_L\xi_L^2\,.
\ee
If there is a sufficient force available (represented by $\xi_L$), then 
the~gate time $\tau_L$ is limited by the criteria that (i) the ions are not
displaced too far ($\bar{x}/d\approx a\xi_L/d\ll 1$), and that (ii) 
the adiabatic conditions following from Eqs.~(\ref{lin10})
and~(\ref{lin12}), 
that is $\tilde{\omega}\tau_L\gg 1$ and $\omega\tau_L\gg 1$, hold.
Typically, the latter is the limiting condition,
and it is sufficient to choose
$\omega\tau=\omega\tau_L\geq5$ because to be precise what we require
in practice is $e^{-(\omega\tau)^2}\to 0$ rather than 
$\omega\tau\gg 1$.
It follows that for a~given value of $\omega\tau$ to achieve 
a desired overall phase the pushing force is smaller 
when the ions are in a linear
trap ($\epsilon=2$) compared to when they are in two separate microtraps
($\epsilon < 2$). The ratio is
\be
\label{lin18}
\frac{\xi_L}{\xi}\approx\sqrt{\frac{3\epsilon}{2}}\,.
\ee
This can be useful to reduce photon
scattering (see below).

The infidelity of a two-qubit phase gate for ions in two separate microtraps
with $\epsilon\ll 1$ was discussed in detail in Ref.~\cite{ions3}. 
The dynamics when we consider general $\epsilon$ was
presented in Sec.~\ref{general}. This permits us to make a~direct 
comparison between the case when the ions are in separate traps
($\epsilon\ll 1$) and in the same trap ($\epsilon=2$), and thus carry over
the results of the previous study in Ref.~\cite{ions3} to the present one.

We assume the state-selective pushing force is the optical dipole force 
produced by an intense laser beam (see the last paragraph in
Sec.~\ref{impl}) in either the travelling-wave (TW)
or the standing-wave (SW) configuration.
There are two main contributions to the total infidelity of the gate:
(i) thermal averaging and (ii) photon scattering.

Two types of thermal averaging take place. These are the averaging
over the dynamic phases $\Theta_{\a\b}$ (${\cal D}$), 
and the averaging over 
the spatial dependence of the pushing force (${\cal S}$). 
We show in Appendix~\ref{A} that
the contribution~${\cal D}$ is of a similar order of magnitude for
ions in separate traps and  in the same trap. We showed in
Ref.~\cite{ions3} that the contribution~${\cal S}$ dominates~${\cal D}$ 
for ions
in separate traps when the waist of the laser beam is $w < d$ (TW), 
or the laser wavelength is $\lambda < d$ (SW). It follows
that the same conditions cause~${\cal S}$ to dominate~${\cal D}$ for ions 
in the same trap.

The averaging over the spatial profile of the force (${\cal S}$) 
is associated with the
thermal motion of either ion with respect to the laser field illuminating
it. We assume for the sake of argument that for the thermal oscillations of
two ions in the same trap, the centre-of-mass mode ($\omega$) and 
the breathing mode ($\omega\sqrt{3}$) have the same temperature. 
Part of the random phase associated with thermal motion 
may be cancelled by a common-mode
rejection, but to be cautious here we will ignore that possibility
and therefore we overestimate the infidelity.
In this case the infidelity associated with process~${\cal S}$ 
is approximately the same as calculated in Ref.~\cite{ions3} for given values
of trap frequency $\omega$, temperature $T$ and laser parameters.
We assume the spin-echo sequence ($\pi$~pulses) is used, 
so we have two gates
each with $\vartheta = \pi/2$, and we add their two infidelities.
For travelling wave excitation with Gaussian laser beams
(each ion illuminated by a separate beam) the resulting infidelity is \cite{eq1}
\be
\label{infid2}
{\cal P}_{\rm TW}&\approx&
\frac{\pi}{3}\left(\frac{3\pi k_BT}{\hbar\omega}\right)
\left(\frac{a}{w}\right)^2
\left(
\frac{2x_0}{w}-\frac{w}{2x_0}
\right)^2\nonumber\\[1mm]
& &
+\frac{2}{9}\left(\frac{3\pi k_BT}{\hbar\omega}\right)^2
\left(\frac{a}{w}\right)^4
{\cal Q}(2x_0/w)\,,\nonumber\\[1mm]
\ee
where ${\cal Q}(y)=12y^4-64y^2+89-34/y^2+1/y^4$,
$\omega$ is the trapping frequency, $a=\sqrt{\hbar/m\omega}$,
$w$ is the size of the waist of the laser beam, and $x_0$ is the position
of the ions in the profile of the laser beam.
Placing the ions at $x_0=w/2$ offers a useful improvement in 
the~performance of the gate.

For standing wave excitation (i.e. the pair of ions positioned in an optical
standing wave along $z$) we have~\cite{eq2}
\be
\label{infid3}
{\cal P}_{\rm SW}&\approx&
\frac{\pi^2}{128}
\left\{
1-\exp\left[-16(k_{\a}a)^2\left(\frac{k_BT}{\hbar\omega}\right)\right]
\right\}^2\,,\nonumber\\[1mm]
\ee
where $k_{\a}=(4\pi/\lambda)\sin(\a/2)$, $\alpha$ is the the angle between two
laser beams forming the standing-wave field, 
$z_0$ is the ion position in this field, and
we assumed that the ions are placed at $k_{\a}z_0=\pi/4$
which minimizes the infidelity and simplifies its expression.

The expression for the number of scattered photons $N$
during the gate operation in a linear trap is 
derived in the same way as the result for ions in separate traps in
Ref.~\cite{ions3}, except that now we consider 
$\tau_L=3\sqrt{\pi}/(\omega\xi_L^2\sqrt{2})$, where we assume the 
spin-echo sequence (hence $\vartheta=\pi/2$) and we add the numbers of scattered
photons in the pair of gates.
Treating the electronic transition giving rise to the dipole force
by a two-level atom model, the result for travelling-wave excitation is
\be
\label{scat_trav}
N_{\rm TW}\approx
C'\,\frac{w^6}{x_0^2}\frac{\omega^2}{P}\,e^{2(x_0/w)^2}\,,
\ee
where $P$ is the laser power, and
\be
C'=\frac{\pi^4c}{2\sqrt{2}}\frac{m}{\lambda^3}\,,
\ee
where $c$ is the speed of light, $m$ is the ion mass, and $\lambda$ 
is the laser wavelength. 

For standing-wave excitation we obtain
\be
N_{\rm SW}&\approx&
C''\,w^2\,\frac{\omega^2}{P}\frac{1}{\cos^2(k_{\a}z_0)}\,,
\ee
where \footnote{There is a mistake in Eq.~(100b) in Refs.~\cite{ions3, ions4},
where $\sin(\alpha/2)$ should read $\sin^{-2}(\alpha/2)$.}
\be
C''=
\frac{\pi^2c}{4\sqrt{2}}\frac{m}{\lambda}\,\frac{1}{\sin^2(\a/2) }\,.
\ee

The total infidelity of the phase gate can be expressed as
\be
\label{infid_tot}
{\cal P}_{\rm tot}={\cal P}+N\,,
\ee
assuming ${\cal P},N\ll 1$. 

\begin{figure}[htb]
\includegraphics[width=8cm]{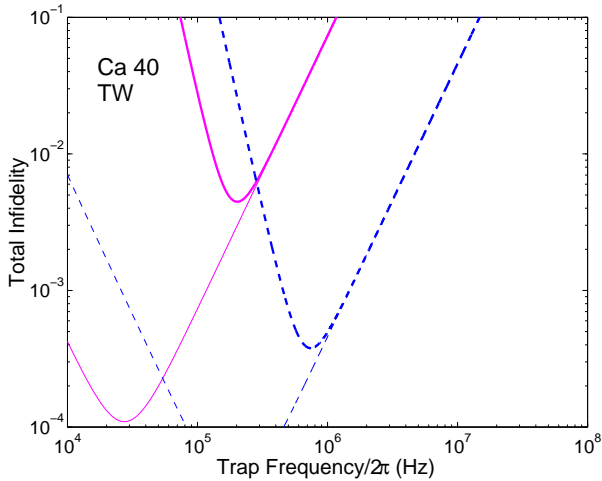}
\caption{The total infidelity of a two-qubit phase gate for two ions in 
a linear trap as a function
of the trap frequency for various parameter values for travelling-wave 
excitation. The solid lines are for $w=4\,\mu$m and $P=10\,$mW, the dashed
lines are for $w=2\,\mu$m and $P=100\,$mW. The thick lines are for the
temperature equal to the Doppler temperature ($T_{\rm dopp}=538\,\mu$K)
associated with the dipole-allowed $4S_{1/2}\leftrightarrow 4P_{1/2}$
transition at $\lambda=397\,$nm in $^{40}\mbox{Ca}^+$. The thin lines are
for $T=\hbar\omega/(k_B\ln2)$. 
The gate time is $\tau_L=5/\omega$.}
\label{ca_trav}
\end{figure}

\begin{figure}[htb]
\includegraphics[width=8cm]{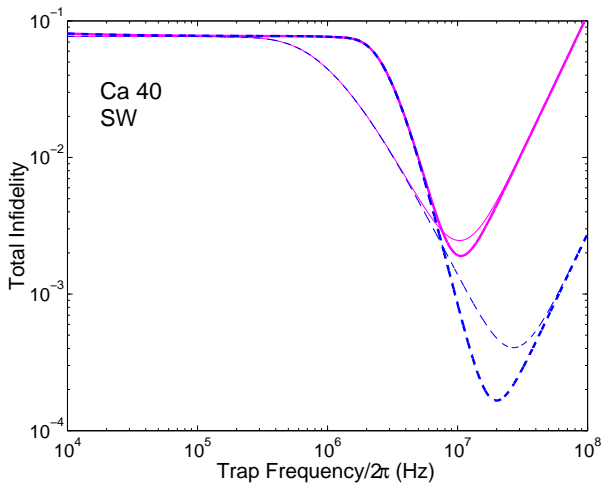}
\caption{The same as in FIG.~\ref{ca_trav} but for standing-wave
excitation, with $\alpha = \pi/2$ and $k_{\alpha}z_0 = \pi/4$.}
\label{ca_stan}
\end{figure}

FIG.~\ref{ca_trav} and FIG.~\ref{ca_stan} 
give the total infidelity for some example parameter choices,
with atomic properties appropriate to the $^{40}$Ca$^+$ ion.  

\section{Conclusion}

In this paper we offered further quantum-gate methods using the
state-selective displacement of trapped atomic ions~\cite{ions1}, and in
this sense this paper complements our previous publication~\cite{ions3}.

We proposed methods to suppress unwanted effects in the evolution,
as an alternative to the spin-echo method. 
We also discussed the application of the gate to logical qubits encoded
in pairs of ions (rather than in single ions) in a decoherence-free subspace.

We showed that it is possible to realize two-qubit gates between non-neighbouring
ions without the need to swap the ions around 
or move quantum information between them. This produced an implementation
of the Toffoli (three-qubit
controlled-NOT) gate with three ions in three separate microtraps, where the
three-qubit gate is five to eight times slower than its two-qubit
counterpart. 

We also showed that the original state-selective displacement method for
a two-qubit phase gate with ions in separate microtraps can be extended to
the case with two ions in the same trap. We analyzed the
anharmonicity of the effective trapping potential, arising from 
the Coulomb interaction potential between ions, and found that its 
effects were small even in the limit of two ions in the same trap.
Therefore the gate retains its attractive features (a good combination
of speed and robustness), whatever the separation of the ion traps.

\begin{acknowledgments}

This work was supported by the EPSRC, ARDA (P-43513-PH-QCO-02107-1), 
and by the Research Training, Development and Human Potential Program QUEST
of the European Union. We would like to acknowledge helpful discussions
with T. Calarco.

\end{acknowledgments}

\appendix
\section{}
\label{A}

We showed in Ref.~\cite{ions4} that in the regime 
$\epsilon\ll 1$ a contribution from thermal averaging to the total infidelity
of a two-qubit phase gate (defined in Eq.~(\ref{infid_tot}))
is given by Eqs.~(\ref{infid2})
and~(\ref{infid3}). In this contribution thermal averaging over the size of
the state-selective force (i.e. over the spatial profile of the force)
dominates averaging over the dynamic phase $\Theta_{\a\b}$ (i.e. over a
discrepancy between random values of $\Theta_{\a\b}$ in a $G$ pulse and
their deterministic values in single-qubit rotations~$S$), where the latter
reads~\footnote{See Eq.~(89) in Ref.~\cite{ions3}.}
\be
\label{fid0}
{\cal P}_{\epsilon\ll 1}=
\left(\frac{3\pi k_BT}{\hbar\omega}\right)^2
\left(\frac{a}{d}\right)^4\,.
\ee

In the regime of interest in this paper ($\epsilon=2$) we need to reconsider
this conclusion on the grounds of anharmonic effects present in the system.
In particular, we need to calculate the contribution to the infidelity from
thermal averaging over the dynamic phases for $\epsilon=2$ and compare the
result to the contribution from averaging over the force profile, which is the
same for any value of~$\epsilon$.

Using a precise analytical expression of the overall phase $\vartheta$ 
in the regime
$\epsilon=2$ we calculate that the infidelity from averaging over the
dynamic phases is
\be
\label{fid1}
{\cal P}_{\epsilon=2}=
\left(\frac{\vartheta_Lk_BT}{3\hbar\omega}\right)^2
\left(\frac{a}{d}\right)^4
\left[2\left(\frac{2}{3\omega\tau}\right)^2+1\right]^2\,.
\ee
When we use the phase condition $\vartheta_L=\pi$, and choose
$\omega\tau=5$, we obtain
\be
\label{fid2}
{\cal P}_{\epsilon=2}\approx
1.07\left(\frac{\pi k_BT}{3\hbar\omega}\right)^2
\left(\frac{a}{d}\right)^4\,.
\ee

Thus we find that the contribution arising from
anharmonicity scales as $(a/d)^4$, and therefore it does not
dominate the infidelity.
Comparing Eqs.~(\ref{fid0}) and (\ref{fid2}), 
the expressions have the same functional form and a similar size.
Both are small compared to Eqs.~(\ref{infid2}) and (\ref{infid3})
when the laser beam waist or the standing wave period
are small compared to the ion separation $d$.


\end{document}